# Oscillatory D'yakonov-Perel' Spin Dynamics in Two Dimensional Electron Gases

W.J.H.Leyland[a,b], R.T.Harley[a], M.Henini[c], D.Taylor[c], A.J.Shields[d], I.Farrer[b] and D.A.Ritchie[b]

[a]School of Physics and Astronomy, University of Southampton SO17 1BJ, UK
[b]Cavendish Laboratory, University of Cambridge, Madingley Road, Cambridge CB3 0HE, UK
[c]School of Physics and Astronomy, University of Nottingham NG7 4RD, UK
[d]Toshiba Research Europe Ltd, Cambridge CB4 4WE, UK

**Abstract**

Optical pump-probe measurements of spin-dynamics at temperatures down to 1.5K are described for a series of (001)-oriented GaAs/AlGaAs quantum well samples containing high mobility two-dimensional electron gases (2DEGs). For well widths ranging from 5 nm to 20 nm and 2DEG sheet densities from $1.75 \times 10^{11} cm^{-2}$ to $3.5 \times 10^{11} cm^{-2}$ the evolution of a small injected spin population is found to be a damped oscillation rather than exponential relaxation, consistent with the quasi-collision-free regime of D'yakonov-Perel spin dynamics. A Monte Carlo simulation method is used to extract the spin-orbit-induced electron spin precession frequency $|\Omega(k_F)|$ and electron momentum scattering time $\tau_p^*$ at the Fermi wavevector. The spin decay time passes through a minimum at a temperature corresponding to the transition from collision-free to collision-dominated regimes and $\tau_p^*$ is found to be close to the ensemble momentum scattering time $\tau_p$ obtained from Hall measurements of electron mobility. The values of $|\Omega(k_F)|$ give the Dresselhaus (BIA) coefficient of spin-orbit interaction as a function of electron confinement energy in the quantum wells and show, qualitatively, the behaviour expected from k.p theory.

## 1) Introduction

In semiconductor media which lack inversion symmetry the spin-orbit interaction causes an electron spin to precess as it propagates through the material. This reorientation provides the drive for the electron spin relaxation mechanism discovered by D'yakonov and Perel' [1] and developed for quantum structures by D'yakonov and Kacharovskii [2]. The frequency and axis of precession, represented by the precession vector $\Omega(\mathbf{k})$, depend on the electron wavevector $\mathbf{k}$ so that strong momentum scattering of the electron produces a succession of small random reorientations of the spin, leading to spin relaxation of a spin-polarised electron population. Many experiments on bulk and quantum well III-V systems have demonstrated that this is the dominant spin-relaxation mechanism in all but p-type material [3]. It has two remarkable features. One is 'motional slowing' which occurs in the strong scattering limit, $|\Omega|\tau_p^* \ll 1$ where $\tau_p^*$ is the momentum scattering time; the rate of spin relaxation is proportional to the average precession angle between scattering events so that increasing momentum scattering actually inhibits spin reorientation and the spin relaxation rate ($\tau_s^{-1}$) is therefore proportional to $\tau_p^*$ [1,3]. For spin component $i$ the relaxation rate is

$$\tau_{s,i}^{-1} = <\Omega_\perp^2> \tau_p^* \qquad (1)$$

where $<\Omega_\perp^2>$ is the average square of the component of the precession vector in the plane perpendicular to *i*. The second is that, in the extreme of weak scattering ($|\Omega|\tau_p^*$ >1), an electron spin may precess through several cycles before scattering and the spin dynamics can become oscillatory rather than relaxational [4]. It turns out that this is not a difficult regime to access. Taking, as an example, a 2DEG in a quantum well at low temperatures, the precession frequency is linear in **k** to first order [5] and for a 10 nm GaAs well with electron density $2\times10^{11}$ cm$^{-2}$ can be estimated using the Dresselhaus (BIA) spin-orbit coupling term to be of order 0.2 radians ps$^{-1}$ at the Fermi wavevector. Then, if it is assumed that the scattering time $\tau_p^*$ is the same as the ensemble momentum scattering time $\tau_p$ which determines the electron mobility, the value of electron mobility required to meet the condition $|\Omega|\tau_p=1$ would be ~ 12 m$^2$V$^{-1}$s$^{-1}$, a modest value quite easily obtainable in modulation-doped samples.

In an earlier publication [6], using ultrafast optical techniques, these two types of behaviour were demonstrated experimentally for a two-dimensional electron gas (2DEG) in a high-quality GaAs/AlGaAs quantum well sample. Oscillatory spin evolution at low temperatures was observed to be replaced by exponential relaxation as the temperature was increased. Before these experiments the equality of $\tau_p^*$ and $\tau_p$ was almost universally assumed [2,3] but the measurements emphasised the importance of electron-electron scattering which had been earlier pointed out theoretically [7]. The electron-electron scattering does not directly affect the mobility but is effective in randomizing the spin precessions of individual electrons; therefore, in general, $\tau_p^* < \tau_p$. More recently [8], we investigated in detail the strong scattering regime ($|\Omega|\tau_p^* <1$) for a complete set of quantum well 2DEGs which demonstrated the motional slowing in a particularly clear form and also provided quantitative verification of the role of electron-electron scattering in setting an intrinsic *lower* limit to the spin-relaxation time. In this paper we describe the spin dynamics for the same set of samples, with well-widths ranging from 5nm to 20 nm, at the lowest available temperature where the 2DEGs are degenerate. The measurements are performed in zero magnetic field and involve optical injection of a small population [9] of electrons at the Fermi level which are spin-polarised perpendicular to the plane. In this situation electron-electron scattering is forbidden by the Pauli exclusion principle and we expect $\tau_p^* = \tau_p$. We find oscillatory evolution of the perpendicular spin component in all samples and use a Monte Carlo simulation method to extract values of frequency $|\Omega(\mathbf{k_F})|$ and of scattering time $\tau_p^*$. The precession frequency provides a first direct measurement of the dependence of the Dresselhaus (BIA) spin-orbit interaction on electron confinement in a quantum well system.

**2) Experimental details**

The samples used for these low temperature experiments (Table 1), with one addition, are the same series of (001)-oriented one-side *n*-modulation doped single GaAs/Al$_{0.35}$Ga$_{0.65}$As quantum wells as used in our earlier high temperature study [8]. The additional sample, NU590, has the same structure as other NU-samples but nominal quantum well width of 5.1 nm. Full descriptions of the structures and of characterization of the samples can be found in ref. [8]; we repeat relevant details in Table 1. The electron confinement energy ($E_{1e}$) and the electron density ($N_S$) were obtained from combined photoluminescence (PL) and photoluminescence excitation (PLE) measurements (figure 1). The ensemble momentum-relaxation time ($\tau_p$) was obtained from Hall mobility measurements. The remaining columns, precession frequency at the Fermi energy ($|\Omega(k_F)|$) and single-electron momentum-relaxation time ($\tau_p^*$), are obtained from the low temperature spin evolution measurements to be described.

The spin-dynamics of the 2DEGs were investigated using the same picosecond-resolution polarized pump-probe reflection technique described in detail in ref.[8]. In this case the pump beam intensity was typically 0.5 mW focused to a 60 micron diameter spot giving an estimated photoexcited spin-polarized electron density $5 \times 10^9$ cm$^{-2}$, very much less than the unpolarised electron concentration in the 2DEGs; the probe power density was 25% of the pump. The technique measures the time-evolution of the probe polarization rotation, $\Delta\theta$, which is proportional to the component of spin-polarisation along the growth axis, $<S_z>(t)$. For most of the measurements the samples were mounted in a liquid helium flow cryostat in which cold gas surrounded the sample allowing temperatures down to 5K to be investigated. Some measurements were made with samples immersed in superfluid helium at 1.5K; the values of $|\Omega(k_F)|$ and of $\tau_p^*$ were found to be unchanged within experimental uncertainty between 1.5K and 5K except for $\tau_p^*$ in T539, the sample with the widest quantum well and highest mobility (see table 1).

**3) Results**

Figure 1 shows an example of the PL and PLE spectra at 5K [8], in this case for sample NU211 with, inset, a schematic diagram indicating the n=1 conduction and valence sub-band structure; arrows indicate the peak of the PL spectrum and the onset of the PLE spectrum. By tuning the laser to the onset of PLE it is possible to inject electrons at the Fermi energy of the 2DEG. Figure 2 shows a typical $\Delta\theta$ signal for each sample with the laser tuned to a photon energy close to the PLE onset. In each case the signal is a rapidly damped oscillation with frequency increasing as the width of the quantum well containing the 2DEG is reduced. The solid curves are fits of an exponentially decaying cosine function

$$\Delta\theta = A\, exp(-t/t_0)\, cos(\omega t) \qquad (2)$$

where $A$, $t_0$ and $\omega$ are fitting parameters. As described below, we extract the precession frequency at the Fermi wavevector, $|\Omega(k_F)|$, and the corresponding single-electron momentum relaxation time, $\tau_p^*$, (see Table 1) from these fits. Note that the parameter $t_0$ is the decay constant of the spin component $<S_z>(t)$, the counterpart, in the low-temperature, oscillatory regime, of the spin relaxation time $\tau_s$ in the high temperature, exponential regime.

The observed signals were found to be a strong function of the wavelength of the laser in the region of the PLE onset. The behaviour was qualitatively the same for all the samples and is illustrated with a contour plot in figure 3 for sample NU211. Over a range of photon energies from 1.552eV to 1.559eV $\Delta\theta$ has the type of oscillatory behaviour shown in figure 2 but the amplitude is varying and the phase of oscillations reverses abruptly near 1.555eV, near the peak of the PLE spectrum. We do not analyse this energy-dependence of the signals in detail here [10] but note, as shown in Figure 4a and 4b, that the extracted values of precession frequency and scattering time remain constant across the investigated energy range. Figure 4 also indicates the PLE spectrum (linear intensity scale) over the same range of photon energies. The spectral resolution of both the time evolution and PLE measurements is ~0.5 meV, determined by the pulse duration of the laser. Also at 5K, thermal smearing of the PLE onset will be determined by $k_B T$ ~0.5 meV.

We now discuss interpretation of the oscillatory signals and how we relate the frequency and damping of the experimental data (curves in figure 2) to the electron spin precession frequency $|\Omega(k_F)|$ and momentum scattering time $\tau_p^*$. The optically injected electrons have wavevectors distributed uniformly round the Fermi surface of the 2DEG. Due to the anisotropy of the precession vector [5,6], this will result in a distribution of precession frequencies and therefore to dephasing of the $\Delta\theta$ signal from the injected population even in the absence of momentum scattering. We showed in an earlier publication [6], however, that this effect would lead to a dephasing time of order hundreds of picoseconds and so is far too weak to explain the observed decay of the oscillations. This means that the observed damping must be caused primarily by momentum scattering and, for the purpose of analysing our data, we can ignore the anisotropy of $\Omega(\mathbf{k_F})$. To simulate this situation we have performed a series of Monte Carlo calculations of $<S_z>(t)$ [6] in which we assume an arbitrary population of $10^5$ spin-up electrons injected at the Fermi energy at $t=0$ with an isotropic distribution of in-plane wave vectors. The spins precess at a common frequency $\Omega$ which may be adjusted in the simulation and undergo random changes of wave vector due to elastic scattering with rate $\tau^{-1}$ which is also adjustable. The axis of precession of each electron is as specified by the Dresselhaus or BIA form for the vector $\Omega(\mathbf{k_F})$ [5] and is thus changed by scattering events. Rather than fit the Monte Carlo simulation directly to the experimental data, we generated graphs of $\Omega/\omega$ and $\Omega\tau$ vs $\omega t_0$ by fitting eq. 2 to simulations performed for a range of parameter values. For each pair of values $\omega$ and $t_0$ from a measured spin evolution we could then read off the values of $|\Omega(k_F)|$ and $\tau_p^*$. Typically, for the evolutions shown in figure 2, this process gave values of $\Omega(k_F)$ up to 25% greater than $\omega$ for the most heavily damped case (NU 590, figure 2) and $\tau_p^* \sim t_0/1.9$ in all cases. The results are given in table 1 and in figures 4, 5 and 6.

**4) Discussion**

Theoretically for (001)-oriented quantum wells the precession vector lies in the plane of the well and has components due to the Dresselhaus (BIA) and Rashba (SIA) spin-orbit interaction terms [5,11,12]. Taken to third order in $k$, the magnitude of the BIA term is

$$|\Omega_{BIA}| = \frac{2\gamma}{\hbar} <k_z^2> k_F \left\{ 1 - \left( \frac{4E_F}{E_{1e}} - \frac{E_F^2}{E_{1e}^2} \right) \sin^2\phi \cos^2\phi \right\}^{0.5} \quad (3)$$

where $z$ is the growth direction and $\phi$ is the angle between $\mathbf{k_F}$ and the (100) direction. $<k_z^2>$ is related to the electron confinement energy by $E_{1e} = \hbar^2 <k_z^2>/2m^*$ and $k_F$ to the Fermi energy by $E_F = \hbar^2 k_F^2/2m^*$ where $m^*$ is the electron effective mass. $\gamma$ is the BIA coefficient which we discuss in more detail below. For the present samples the second term in eq.4 is small (in the worst case (T539) it contributes only 12%) making $|\Omega_{BIA}|$ essentially isotropic. The magnitude of the SIA term is

$$|\Omega_{SIA}| = \alpha \frac{e}{\hbar} F k_F \quad (4)$$

where $\alpha$ is the Rashba coefficient and $F$ is the effective built-in electric field in the sample. The BIA and SIA terms have different wavevector dependence so that, in principle, they will interfere to produce anisotropy of the precession frequency [12,13]. However we expect the SIA term to be unimportant in these samples. Firstly we recently measured $\alpha = 0.083 \pm 0.005$ nm$^2$ in a 7.5 nm GaAs/AlGaAs quantum well [14], confirming k.p theory calculations. Secondly we estimate that in these structures the built-in field is $\leq 1.5 \times 10^6$ Vm$^{-1}$ along the growth axis, associated with band bending and band offsets [5,6]. Using these values in eq. 4 we find the SIA term can be neglected in comparison with the experimental uncertainty in all cases. Thus in all our samples, under the conditions of the measurements the precession frequency is determined to within a few percent by the BIA coupling and is essentially isotropic. Eq. 3 reduces to

$$|\Omega(k_F)| = \frac{2|\gamma|}{\hbar} <k_z^2> k_F = \frac{2|\gamma|}{\hbar}\left(\frac{2m^*}{\hbar^2}\right)^{\frac{3}{2}} E_{1e}(E_F)^{0.5} \qquad (5)$$

In figure 5 we plot the BIA coefficient $|\gamma|$ from our measurements against confinement energy $E_{1e}$. At zero confinement the extrapolated value is $|\gamma| \sim 12$ eVÅ$^3$ and there is a clear downward trend with increasing $E_{1e}$.

The main previous experimental determination of $\gamma$ in quantum wells was by Raman scattering in several 18 nm GaAs quantum well samples [12,13], similar to our sample T539 with $E_{1e} \sim 10$ meV. Two values were obtained $\gamma = -16\pm3$ eVÅ$^3$ [12] and $\gamma = -11$ eVÅ$^3$ [13] which are both consistent with our measured value of $|\gamma|$ for low confinement energy. However a direct comparison of the values is not appropriate because the 2DEG concentrations in the Raman samples were much higher than in our samples (6.5 x10$^{11}$ cm$^{-2}$ to 14.0x10$^{11}$ cm$^{-2}$) resulting in stronger band-bending such that the confining potential was essentially that of a heterojunction rather than a quantum well. There was also a significant Rashba (SIA) contribution to the spin splitting which enabled determination of the sign of $\gamma$.

Theoretical estimations of the BIA coupling coefficient have been reviewed by Winkler [5]. For bulk GaAs 14-band k.p perturbation theory in third order gives $|\gamma| = 27.58$ eVÅ$^3$ and including remote bands in fourth order $|\gamma| = 19.55$ eVÅ$^3$. For GaAs/AlGaAs quantum wells $|\gamma|$ is expected to be reduced and to fall with confinement energy because of associated increases in the energy denominators appearing in k.p theory and, to a lesser extent, because of wavefunction penetration into the barriers where the value of $|\gamma|$ is lower than in the well. Thus it seems likely that a detailed k.p calculation for these samples will account for our observed values of $|\gamma|$.

Theoretically, we expect that the extracted values of $\tau_p^*$ will be close to $\tau_p$ (see Table 1) because we are dealing with electrons injected close to the Fermi level in almost fully-degenerate 2DEGs where, in principle, electron-electron scattering is forbidden by the Pauli principle [15]. Their ratio is very close to unity for the widest quantum well (T539) but apparently tends to increase with confinement energy. However given the experimental uncertainties, it seems unlikely that these departures by a factor two to five from unity represent a lack of understanding of the physics of the situation. The electron-electron scattering becomes more than an order of magnitude stronger than any other mechanism at the Fermi temperature $E_F/k_B$ [8] making $\tau_p/\tau_p^*$ extremely sensitive to a small departure from full degeneracy.

Figure 6 illustrates the relaxation times across the transition from the low-temperature quasi-collision-free regime to the high-temperature collision-dominated regime of spin dynamics in two of the samples (a) T539 and (b) NU211. In the high temperature regime exponential spin relaxation is observed (as described in [8]) with the relaxation time given by the D'yakonov-Perel' formula, equation 1. The rapid increase with temperature is caused predominantly by the increasing contribution of electron-electron scattering to the scattering rate $\tau_p^{*-1}$. There is only a relatively small increase of $<\Omega_\perp^2>$. In the low temperature regime, where the spin evolution is observed to be oscillatory, the solid points in figure 6 are values of $(\Omega(k_F)^2 \tau_p^*)^{-1}$ calculated using the quantities extracted from the oscillatory decays. They join smoothly with the high temperature data but in this regime they do not provide a measure of spin relaxation. In the quasi collision-free regime the decay of the spin polarization is given by the decay constant of the oscillatory signal, $t_0$ in eq. 2 (open triangles). The horizontal arrows in figure 6 indicate the values of $|\Omega(k_F)|^{-1}$ for the two samples, corresponding to the condition $|\Omega(k_F)|\tau_p^* =1$; it can be seen that the spin decay time passes through a minimum near this condition.

## 5) Conclusions

We have investigated the low temperature spin dynamics of a set of high mobility 2DEGs confined in (001)-oriented GaAs/AlGaAs quantum wells of different width. In each sample we observe oscillatory spin evolution at 5K or below from which we obtain the spin-orbit splitting of the n=1 conduction sub-band as a function of electron confinement energy. The electron concentrations in these 2DEGs are low compared to samples used in previous studies [5,12,13], consequently built-in electric fields are sufficiently low that the splitting is determined to good accuracy by the Dresselhaus (BIA) spin-orbit interaction term alone. We are thus able to extract, for the first time, the coupling coefficient as a function of electron confinement energy. From the observed spin evolution we have also obtained the spin decay time and the electron momentum scattering time. The spin decay time is found to pass through a minimum at a temperature corresponding to the transition from collision-free to collision-dominated regimes of spin dynamics. This is a further illustration of the motional slowing which is a characteristic of the D'yakonov-Perel' spin relaxation mechanism. The electron momentum relaxation time, $\tau_p^*$, is expected to be equal to the ensemble momentum relaxation time, $\tau_p$, in a fully degenerate 2DEG at the Fermi energy. In our measurements at the lowest temperatures we find that they are nearly equal in the sample with widest quantum well but that $\tau_p^*$ becomes significantly shorter than $\tau_p$ for narrower wells. This is consistent with small departures from full degeneracy since we know from our high-temperature measurements and analysis in ref. [8] that $\tau_p^*$ is very strongly influenced by electron-electron scattering.

**Table 1.** Sample parameters

| Sample No. | Electron confinement energy $E_{1e}$ (meV) | Nominal well width $L_z$ (nm) | Electron density $N_S$ (cm$^{-3}$) | Fermi energy $E_F$ (meV) | Ensemble momentum-relaxation time $\tau_p$ at 5K (ps) | $|\Omega(k_F)|$ (ra ps$^{-1}$) | Momentum relaxation time $\tau_p^*$ (ps) 1.5K | 5 K |
|---|---|---|---|---|---|---|---|---|
| T539 | 10.2 | 20.0 | 1.75 10$^{11}$ | 6.2 | 27 | 0.063±0.006 | 22±3 | 12.5±2.0 |
| T315 | 49.8 | 10.0 | 1.86 10$^{11}$ | 6.8 | 10 | 0.19± 0.01 | 6.0±0.2 | 6.0±0.2 |
| NU211 | 32.8 | 10.2 | 3.10 10$^{11}$ | 11.1 | 13 | 0.22±0.01 | - | 6.4±0.9 |
| NU535 | 58.5 | 6.8 | 3.30 10$^{11}$ | 11.9 | 13 | 0.29±0.02 | 5.0±0.9 | 5.1±0.9 |
| NU590 | 95.1 | 5.1 | 3.50 10$^{11}$ | 12.5 | 13 | 0.41±0.08 | - | 2.5±1.1 |

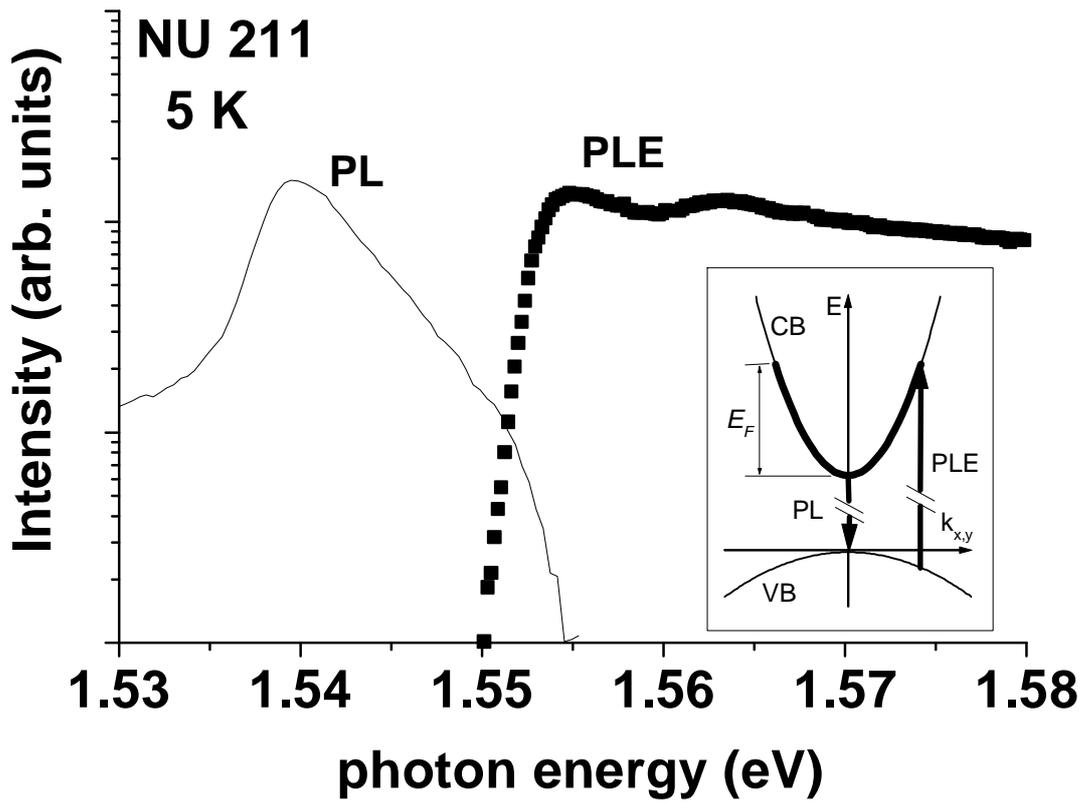

**Figure 1** Typical photoluminescence (PL) and photoluminescence excitation (PLE) spectra of one sample (see ref. 8 for other samples) from which electron density and confinement energy were obtained. Inset illustrates the n=1 subbands of the quantum well with arrows indicating PLE onset and PL peak. Spin dynamic measurements were made at photon energies across the PLE onset.

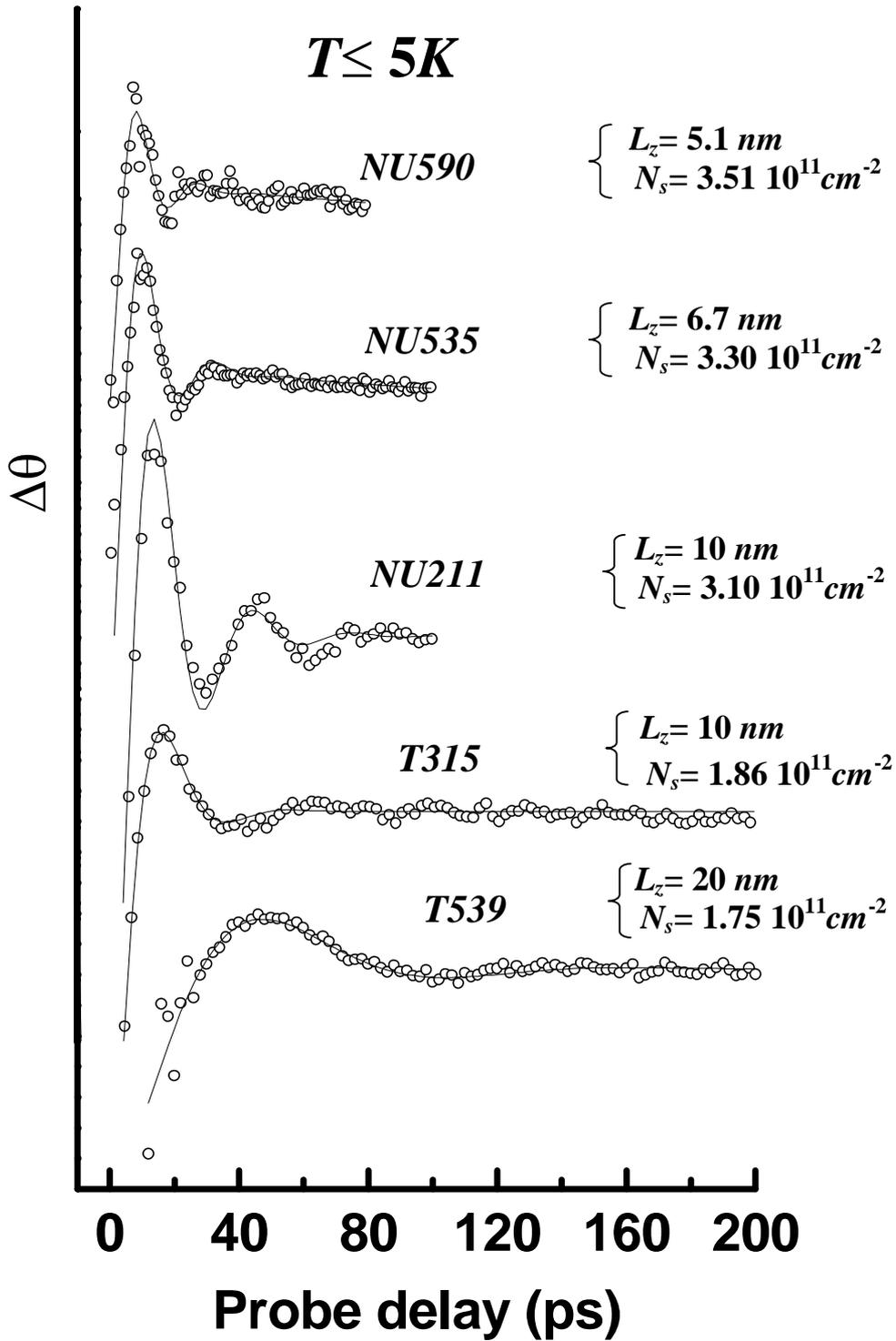

**Figure 2** $\Delta\theta$ signals at the lowest temperature studied for each sample, showing the time-evolution of $<S_z>(t)$ (zeros offset for clarity). Solid curves are fits of eq. 2 to the experimental points.

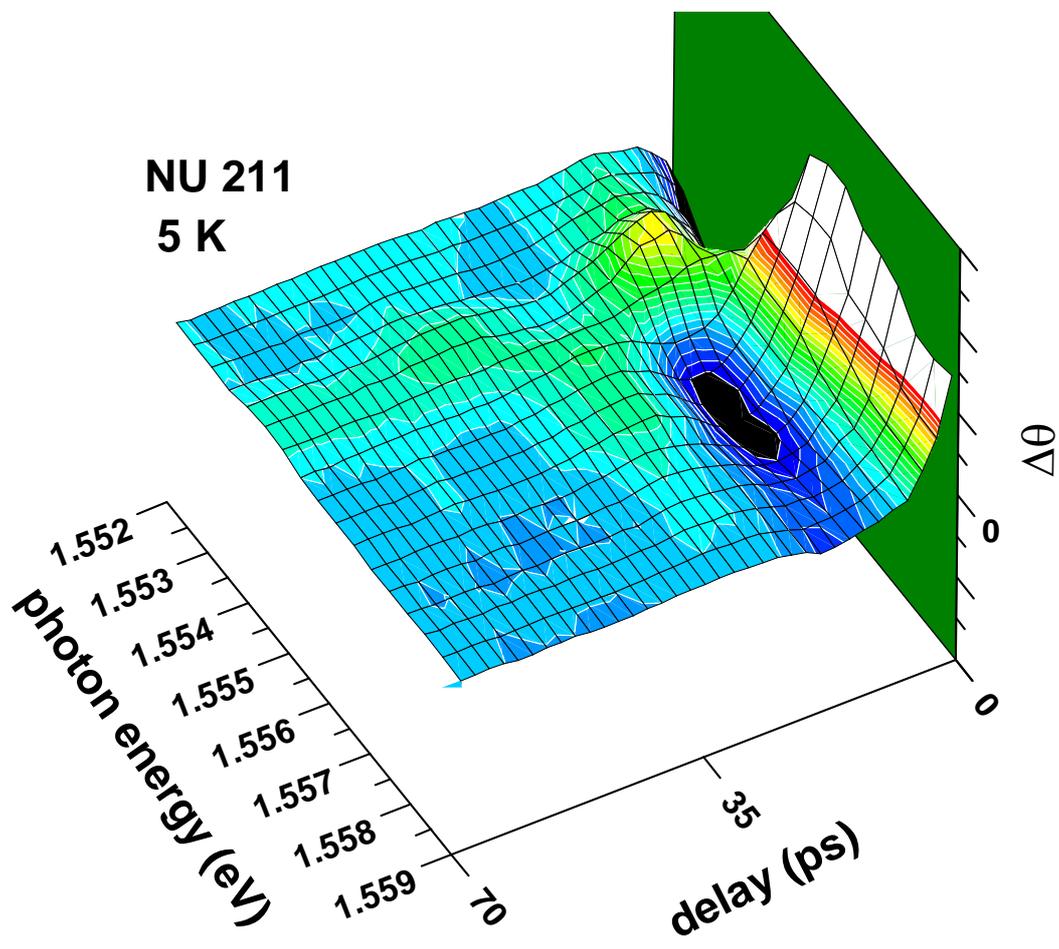

**Figure 3** (Color online) Contour plot of $\Delta\theta$ signals for different photon energies in sample NU211. Note that the signal is oscillatory for all energies but reverses sign close to 1.555eV corresponding to the peak of PLE (see figs. 1 and 4).

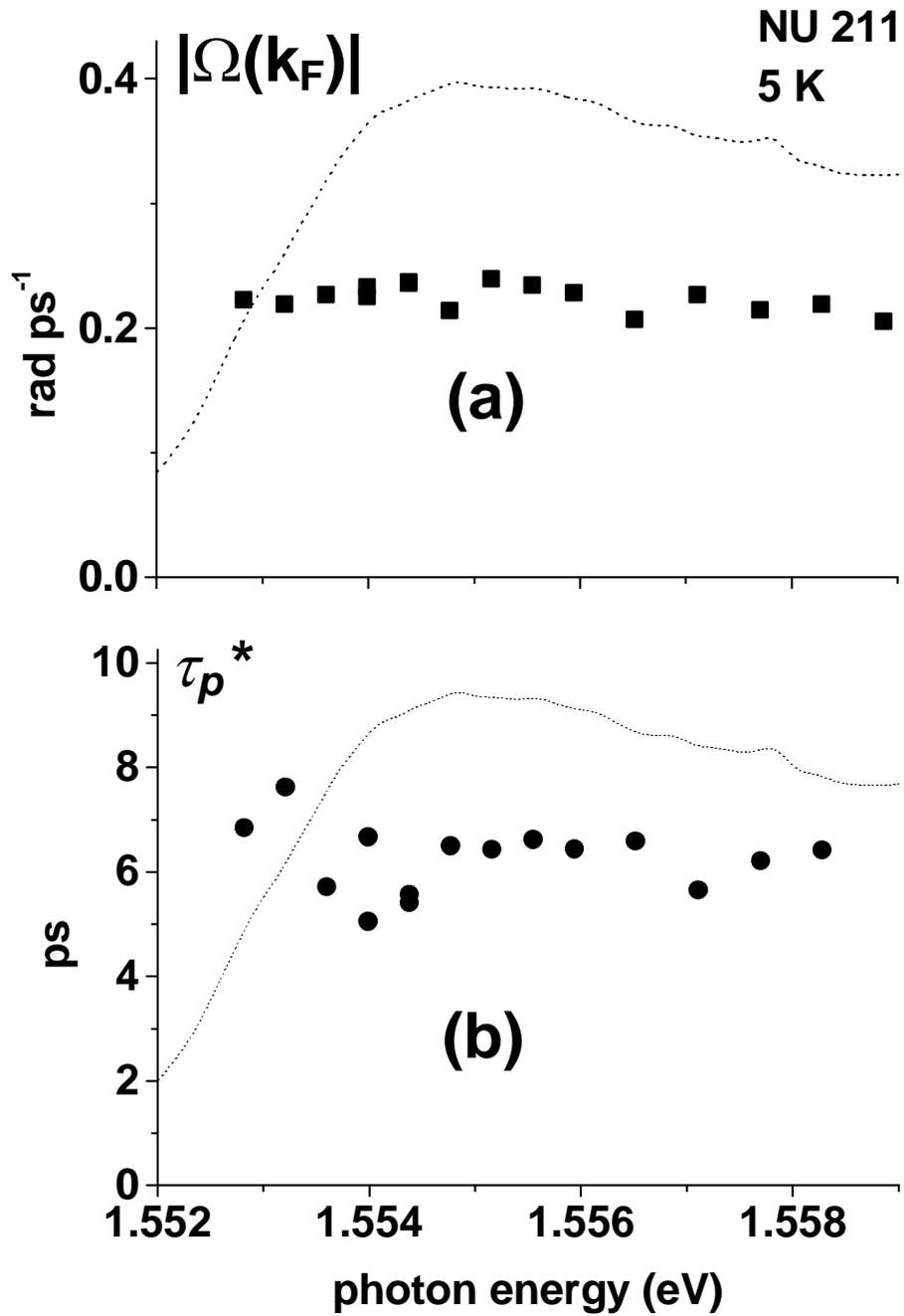

**Figure 4**  Experimental values at 5K of (a) precession frequency ($|\Omega(k_F)|$) and (b) of electron momentum scattering time ($\tau_p^*$) for different photon energies across the PLE onset (dotted, linear intensity scale) for sample NU211, extracted from data of figure 3 using Monte Carlo simulation method described in the text.

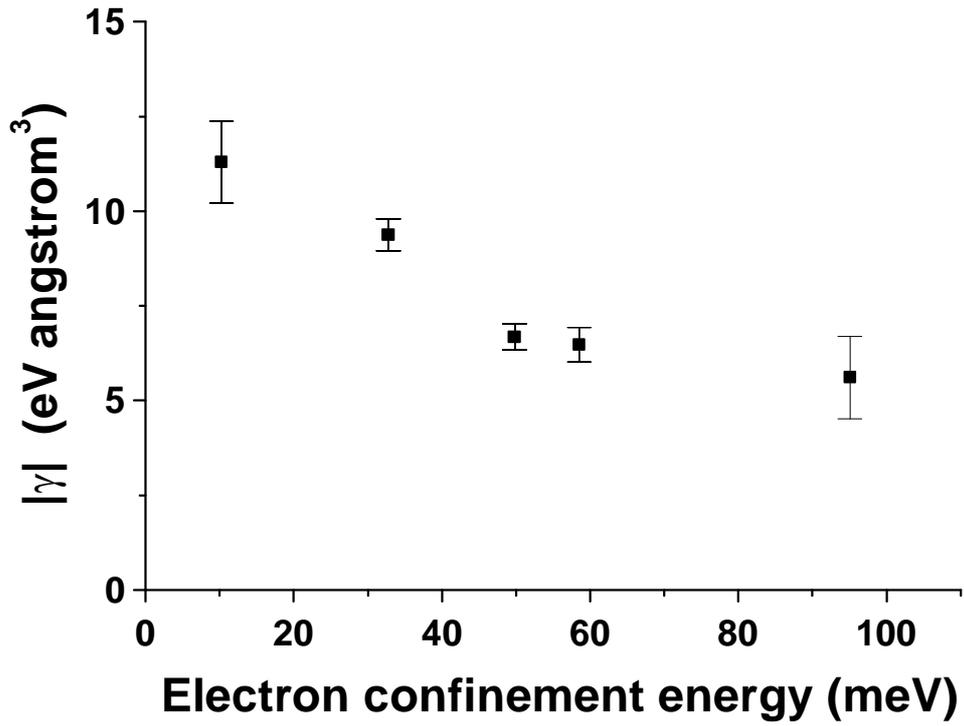

**Figure 5**     Experimental values of Dresselhaus (BIA) spin-orbit coupling coefficient $|\gamma|$ for different electron confinement energies calculated from the measured precession frequencies in the different 2DEG samples. The sign of the coefficient is not determined by these measurements.

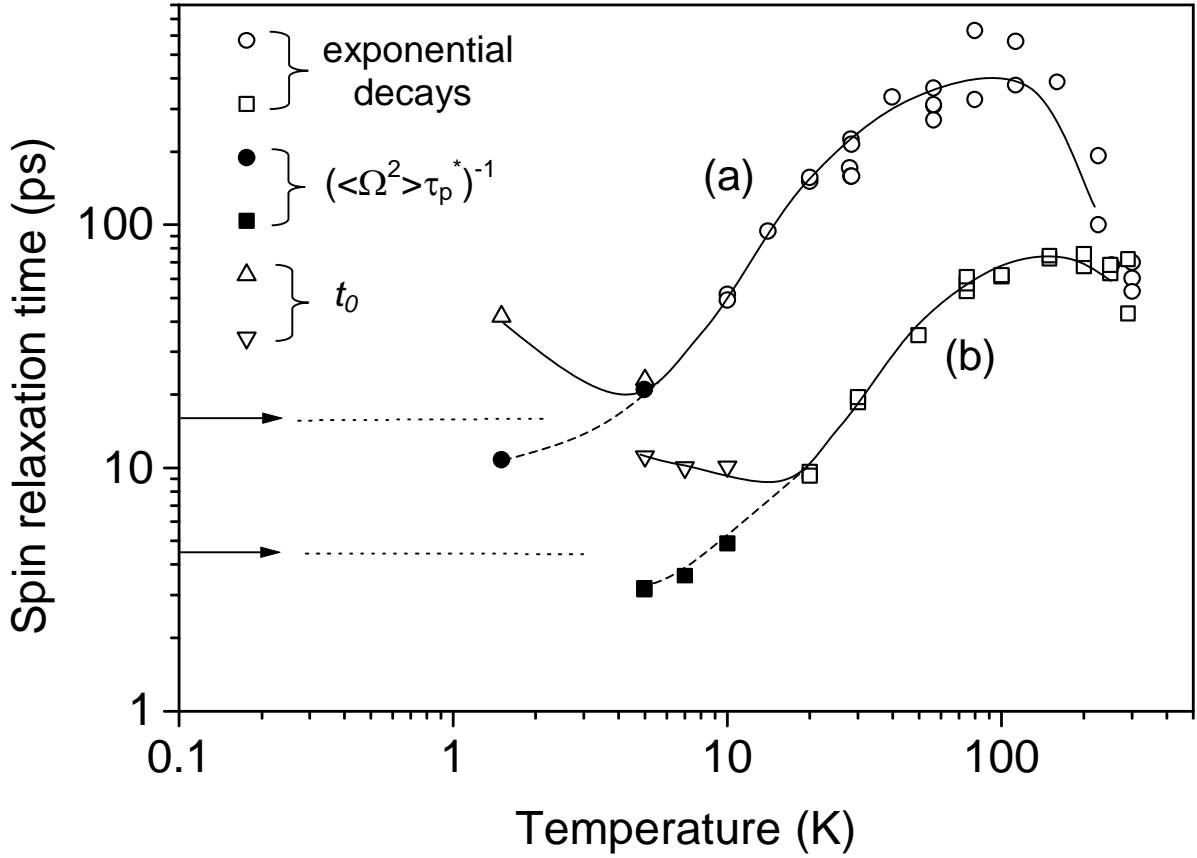

**Figure 6** Measured spin relaxation times for (a) sample T539 and (b) sample NU211. The curves are guides for the eye. Open circles and squares are spin relaxation times measured in the high temperature regime where spin-evolution is exponential (see ref 8). Solid symbols are values of $(|\Omega(k_F)|^2 \tau_p^*)^{-1}$ obtained from analysis of the spin evolution in the low-temperature oscillatory regime, as described in the text. Open triangles are the decay time constant of the oscillatory spin evolution, $t_0$, obtained by fitting eq. 2 to the experimental data. Arrows indicate the value of $|\Omega(k_F)|^{-1}$ for each sample, corresponding to the condition $|\Omega(k_F)|\tau_p^* = 1$.